\documentclass[structabstract]{aa}
\usepackage{graphicx}
\usepackage{amsmath}
\usepackage{epsfig} 
\usepackage{txfonts,ulem}
\usepackage{xcolor}
\usepackage{url}
\usepackage{natbib}
\usepackage{longtable}
\bibpunct[]{(}{)}{;}{a}{}{,}
\usepackage[caption=false]{subfig}
\usepackage{lipsum}

\newcommand{\sepspeed}{$95 \pm 13$~m\,s$^{-1}$}

\newcommand{\sepspeedoneday}{$         178 \pm           22$~m\,s$^{-1}$}

\newcommand{\avesepspeed}{$229\pm 11$~ms$^{-1}$}

\newcommand{\medianflux}{$4.6\times 10^{21}$~Mx}   
\newcommand{\medianfluxsetone}{$4.7\times 10^{21}$~Mx}   

\begin{document}

\title{
Average motion of emerging solar active region polarities \\
I: Two phases of emergence
}
\titlerunning{Two phases of solar active region emergence}

\author{
H.~Schunker \inst{\ref{inst1}}
\and
A.~C. Birch \inst{\ref{inst1}}
\and
R.~H. Cameron\inst{\ref{inst1}}
\and
D.~C. Braun  \inst{\ref{inst3}} 
\and
L.~Gizon \inst{\ref{inst1},\ref{inst2}}
\and
R.~B. Burston \inst{\ref{inst1}}
}

\institute{
Max-Planck-Institut f\"{u}r Sonnensystemforschung, 37077  G\"{o}ttingen, Germany \label{inst1}\\
\email{schunker@mps.mpg.de}
\and
Georg-August-Universit\"{a}t G\"{o}ttingen, Institut f\"{u}r Astrophysik, Friedrich-Hund-Platz 1, 37077   G\"{o}ttingen, Germany\label{inst2}\\
\and
NorthWest Research Associates, 3380 Mitchell Ln, Boulder, CO 80301, USA  
\label{inst3}
}

\date{Received $\langle$date$\rangle$ / Accepted $\langle$date$\rangle$}

\abstract
{}
{Our goal is to constrain models of active region formation by tracking the average motion of active region polarity pairs as they emerge onto the surface.
}
{
We measured the motion of the two main opposite polarities in 153 emerging active regions (EARs) using line-of-sight magnetic field observations from the Solar Dynamics Observatory  Helioseismic Emerging Active Region (SDO/HEAR) survey (Schunker \textit{et al.} 2016). 
We first measured the position of each of the polarities eight hours after emergence, when they could be clearly identified, using a feature recognition method.
We then tracked their location forwards and backwards in time.
}
{
We find that, on average, the polarities emerge with an east-west orientation and in the first 0.1~days the separation speed between the polarities increases. 
At about 0.1~days after emergence, the average separation speed reaches a peak value of \avesepspeed, and then stars to decrease, and about 2.5~days after emergence the polarities stop separating.
We also find that the separation and the separation speed in the east-west direction are systematically larger for active regions with higher flux.
The scatter in the location of the polarities increases from about $5$~Mm at the time of emergence to about $15$~Mm at two days after emergence.
}
{
Our results reveal two phases of the emergence process defined by the rate of change of the separation speed as the polarities move apart. 
Phase 1 begins when the opposite polarity pairs first appear at the surface, with an east-west alignment and an increasing separation speed.
We define Phase 2 to begin when the separation speed starts to decrease,  and ends when the polarities have stopped separating.
This is consistent with the picture of Chen, Rempel, \& Fan (2017): the peak of a flux tube breaks through the surface during Phase 1. During Phase 2 the magnetic field lines are straightened by magnetic tension, so that the polarities continue to move apart, until they eventually lie directly above their anchored subsurface footpoints.
The scatter in the location of the polarities is consistent with the length and time scales of supergranulation, supporting the idea that convection buffets the polarities as they separate. 
}
\maketitle


\section{Introduction}\label{sect:intro}

Solar activity is driven by magnetic fields resulting from a dynamo operating inside the Sun. The processes that generate the magnetic field are largely hidden from view and need to be inferred from what we see at the surface. Therefore,  observations of magnetic flux emerging through the solar surface are a key element to further constrain the physics of the dynamo.


Recent simulations by \cite{Chenetal2017} coupled rising flux tubes from global dynamo simulations  with Cartesian simulations of near-surface convection, naturally forming sunspots with penumbrae. Their simulations show that after coherent flux is first observed at the surface, it takes a few days for most of the flux to emerge.  The final location of the polarities at the surface lie directly above the foot-points of the flux tube at the bottom boundary, $32$~Mm below the surface.

Simple active regions consist of a pair of opposite polarities that grow in size and magnetic flux, separate and develop a tilt angle \cite[for a summary of observed properties see][]{vDGGreen2015}. They have typical lifetimes of days (lower flux regions) to weeks (higher flux regions).
The observed motion of the polarities at the surface is due to a combination of magnetic tension, drag force and advection. 
Magnetic tension is proportional to the square of the magnetic field and the curvature; the drag force is proportional to the cross-sectional area of the polarity, the square of the speed of the polarity relative to the surrounding plasma, the plasma density and Reynold's number; and the advection is proportional to the local flows. Understanding the motion of the polarities in relation to these quantities will help us to better constrain  models of emerging active regions.




High-cadence, full-disk observation monitoring campaigns such as the Michelson Doppler Imager onboard the Solar and Heliospheric Observatory  \citep[SOHO/MDI; ][]{SOHO1995} and the Helioseismic and Magnetic Imager onboard the Solar Dynamics Observatory \citep[SDO/HMI; ][]{SDO2012} have provided a wealth of data. These observations make it possible to perform statistical analyses tracing the evolution of active regions with higher and spatial and temporal resolution \cite[e.g.][]{KosovichevStenflo2008,McClintockNorton2016}.

In addition, these instruments make it possible to use helioseismology to measure the plasma flows during the emergence process. 
A statistical analysis of the relationship between surface flows and the individual motion of the polarities is necessary to gain an understanding of the physics of the emergence process.

In this paper we use the Solar Dynamics Observatory Helioseismic Emerging Active Region (SDO/HEAR) survey \citep{Schunkeretal2016} to study the surface motions of the leading and following polarities (the leading polarity is in the Westward, or prograde, direction from the following polarity). 
The SDO/HEAR survey is ideal to study the statistical properties of the individual motion of the polarities and their associated flows because the active regions emerge into relatively quiet regions on the solar surface. The dataset consists of white light, magnetic field and  surface velocity observations.
 The results presented in this paper provide new, statistically significant constraints for models of emerging active regions.

In Sect.~\ref{sect:method} we describe the feature identification method we use to determine the location of the polarities from line-of-sight magnetograms.
In Sect.~\ref{sect:bipos} we show the average motion of the leading and following polarities independently, up to two days after emergence.
In Sect.~\ref{sect:sepflux}  we then show the average separation between the polarities as a function of time and maximum flux. We also discuss the scatter in the position of, and separation between, the polarities.
We present a qualitative picture of flux emergence that is consistent with our results in Sect.~\ref{sect:phases}.

\section{Measuring the location of the magnetic polarities}\label{sect:method}

First, we briefly describe the data we use from the SDO/HEAR survey  \citep{Schunkeretal2016}. We have extended the survey to include an additional 77 emerging active regions, making a total of 182 EARs that were observed by SDO/HMI between May 2010 and July 2014 (see Appendix~\ref{app1}). 

For each EAR, the full-disk SOHO/MDI line-of-sight magnetic field observations were Postel projected onto maps centred on the active region and tracked at the Carrington rotation rate with a cadence of 45~seconds.
The tracked maps are stored as a time-series of datacubes of length  6.825~hours  (547 frames), with the datacubes having a cadence of  5.3375~hours (320.25~minutes, 427 frames), overlapping by 90~minutes. Because we are concerned with the evolution of active regions in the first $2\,-\,4$~days, in this paper we averaged the line-of-sight magnetogram maps over each 6.825~hour datacube. For each active region we computed the flux-weighted emergence location and shifted the maps so that the emergence location was at centre \citep{Birchetal2016,Schunkeretal2016}.

We define the emergence time, $\tau=0$, as the time when the absolute flux, corrected for line-of-sight projection, reaches 10\% of its maximum value over a 36-hour interval following the first appearance of the sunspot (or group) \cite[for more detail see][]{Schunkeretal2016}. Each datacube is labelled with a time interval, \texttt{TI},  relative to the emergence time.  The time interval \texttt{TI+00} covers the time 61.5~minutes (82 frames) before the emergence time to 281.25 minutes (374 frames) after.  Table~\ref{tab:ti} lists the mid-time relative to the time of emergence for each time interval label.

Hale's law \citep{HaleNicholson1925} states that most high-flux active regions consist of roughly east-west aligned pairs of opposite magnetic polarity, with a preferred sign that is opposite in the northern and southern hemispheres.  Joy's law \citep{Haleetal1919} states that the leading (westward) polarity is typically closer to the equator than the following polarity.  

To account for Hale's law when averaging over EARs in the northern and southern hemispheres, we switched the sign of the line-of-sight magnetic field for the regions in the southern hemisphere and to account for Joy's Law we flipped the maps in the latitudinal direction.  
This allows us to define the coordinate axis of each Postel projected region such that the centre of the map is at the origin, with the $+y$-direction pointing away from the equator, and the $+x$-direction towards solar west (in the prograde direction).  All of the active regions occur during solar cycle 24, and so in our coordinate system the leading polarity is preferentially negative and the following polarity is preferentially positive in the northern hemisphere.

\begin{figure*}
\includegraphics[width=0.9\textwidth]{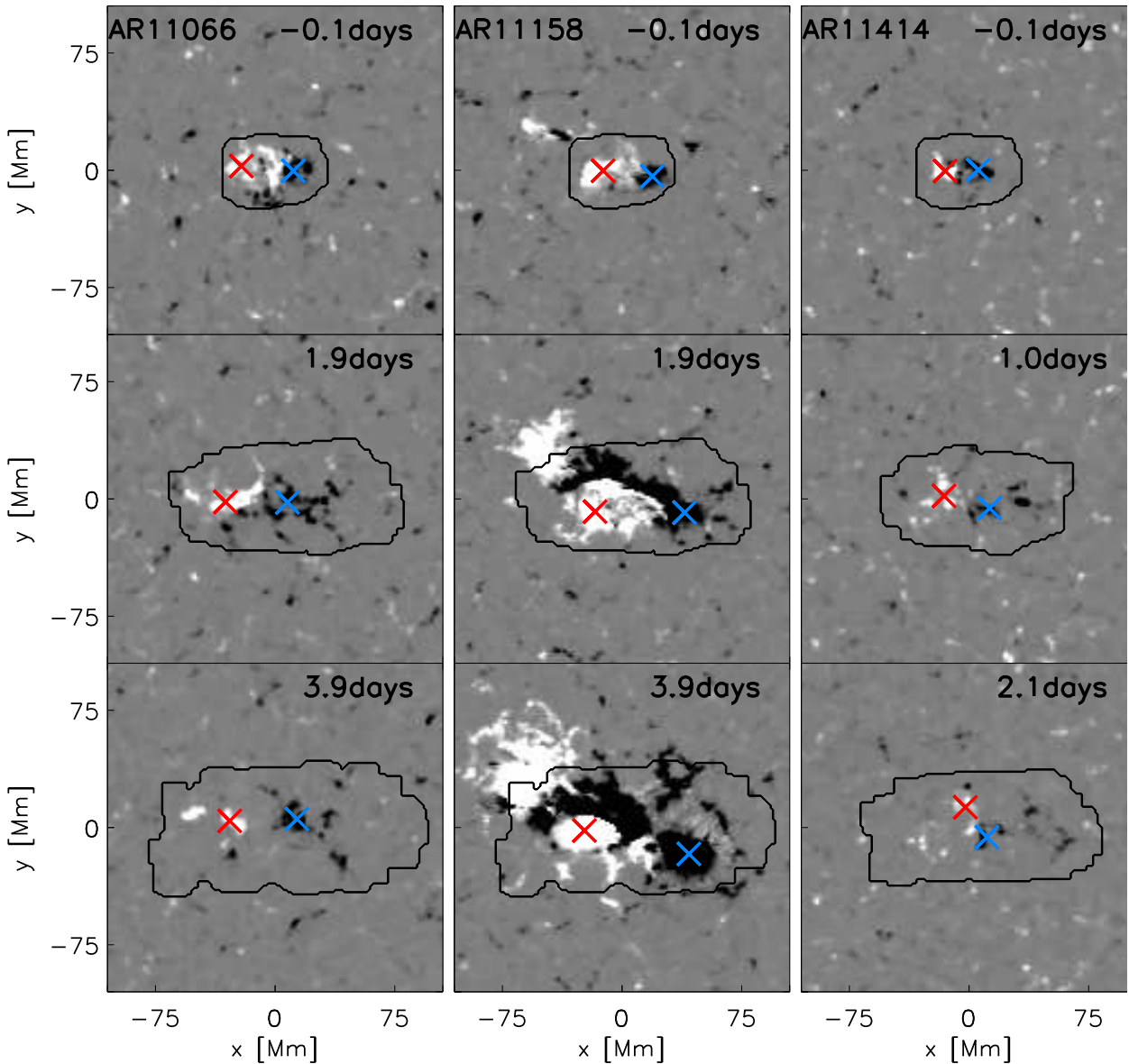} 
\caption{
Position of negative (blue cross) and positive (red cross) polarities on the time-averaged line-of-sight magnetogram for a low-flux active region on the left (AR~11066), a complex, high-flux region in the middle (AR~11158), and a weak short-lived active region (AR~11414) on the right at different times.  The range of the grey-scale is $\pm 100$~G. The black contour indicates the region within which we search for the polarities.
AR~11414 is an example of an active region that was excluded from the analysis in this paper because it is not clear which feature should be defined as the leading polarity. For a full list of excluded regions see Appendix~\ref{app2}.
}
\label{fig:bpmaps}
\end{figure*}

We wanted to study the motion of the leading and following polarities of active regions as they emerge onto the surface of the Sun.
We defined a search area to isolate the bipole associated with the emerging active region from any unrelated surrounding flux.
To do this, we averaged the absolute value of the line-of-sight magnetic field over all 182 active regions at each time interval and smoothed this map with a Gaussian of $\textrm{FWHM}=6.5$~Mm to remove any smaller-scale features. 
Based on a visual inspection, we defined the search area to be limited to all pixels within a radius of 100~Mm from the centre of the map with a value greater than 10~G.  
This resulted in a roughly circular search area at the centre of the map for early time intervals, which increased in size, and became more elliptical with the semi-major axis in the east-west direction in time (see Fig.~\ref{fig:bpmaps}).

At the very beginning of the emergence process the bipoles are small, weak and hard to distinguish from surrounding quiet Sun magnetic fields which do not develop into active regions.
Therefore, we first measured the location of the clearly visible active region polarities at time interval \texttt{TI+02}. We then proceeded backwards (and then forwards) in time sequentially, and selected the identified feature closest to the previously found feature.  This procedure has the advantage that it ignores any newly formed flux from multiple emergences, and tracks the smaller and weaker polarities against background quiet-Sun field.   A drawback of this procedure  is that it does not correctly treat the splitting of features.

Magnetic polarities associated with active regions are generally circular features in line-of-sight magnetograms (see, for example, Fig.~\ref{fig:bpmaps}).  
In order to measure the position of the polarities in each map we used a feature recognition algorithm (\texttt{feature.pro} copyright 1997, John C. Crocker and   David G. Grier) which is designed to determine the centroid position of roughly circular features in an image. 
We prepared our line-of-sight magnetogram maps for the algorithm by first setting all pixels with line-of-sight magnetic field less than $20$~G or that are outside the search area (as described above) to zero. For the negative polarity case, we first switched the sign of the line-of-sight magnetic field and then followed the same procedure.  We selected a threshold value of 20~G to exclude most of the small scale field not associated with the emerging active region. 

At \texttt{TI+02}, we applied the \texttt{feature.pro} algorithm to these prepared maps to first find the maximum within a circular area of diameter 25~pixels (35~Mm) iteratively centred on each pixel in the image array. 
The result of this is a list of unique maxima separated by more than 35~Mm.
We then select the maximum with the largest sum of the prepared line-of-sight magnetic field map within the 35~Mm diameter.
The $x$ and $y$-centroid centred on the maximum within the 35~Mm diameter of the prepared line-of-sight magnetic field map is defined as the location of the polarity.

Moving forward and backwards in time, we repeat the process but instead of choosing the polarity with the largest sum of unsigned flux, we choose the $x$ and $y$-centroid closest to the polarity location in the previous time interval.

Figure~\ref{fig:bpmaps} shows the magnetic field maps and the location of the features we identified for three example emerging active regions. Most active regions  consist of two clearly identified polarities, such as  AR~11066. For the case of complex emergences, as in AR~11158, we show that our feature identification method tracked both the leading $\left( x_{\mathrm l}(\tau),y_{\mathrm l}(\tau) \right)$ and following $\left( x_{\mathrm f}(\tau),y_{\mathrm f}(\tau) \right)$ polarities that we identified also by eye, relative to the emergence location. Time $\tau$ is relative to the emergence time ($\tau=0.1~\mathrm{days} \equiv \texttt{TI+00}$ and  Table~\ref{tab:ti} contains further equivalent time intervals). 

We excluded 29~active regions where it was difficult to track the locations of the polarities correctly or are persistent anti-Hale's law regions (see Appendix~\ref{app2}). Anti-Hale's law active regions are specific anomalies that are necessary to study as a separate set. An instance of an active region where it was difficult to confidently track the location of the polarity, AR11414, is  shown in the right panel of Fig.~\ref{fig:bpmaps}.  The negative (leading) polarity divides after emergence and the feature identification method tracks the smaller, decaying feature. By eye though, it is not clear which feature should be defined as the leading polarity.

Appendix~\ref{app:methcomp} describes three different methods we tried to measure the location of the polarities: the flux-weighted method, the flux-summed method, and the feature identification method. Figure~\ref{fig:expos} supports our decision to use the feature identification method to detect the polarities associated with emergence in more complex active regions. 
We visually determined that the feature identification method successfully tracked the location of the polarities in 153 of the emerging active regions (see Appendix~\ref{app2}).
Our analysis of the motion of active region polarities during emergence is based on these 153 regions.

We emphasise that our method of identifying the polarity is  based on the sign of the magnetic field and \emph{not} on their east-west orientation, since this can change over time. We specifically exclude active regions with sustained anti-Hale orientation, but it is possible that during the first half a day of the emergence some active regions may have an anti-Hale orientation. 
Since this dataset covers only one solar cycle and we are analysing orientations relative to a statistical expectation, we refer to the negative polarity as the leading polarity and the  positive polarity as the following polarity.

\section{Motion of the individual magnetic polarities}\label{sect:bipos}

Since each active region emergence is unique, to understand the dominant physics guiding emerging active regions we propose to measure the average motion of active region polarities.

\begin{figure*}
\centering
\hspace{-1cm}
\includegraphics[width=0.9\textwidth]
{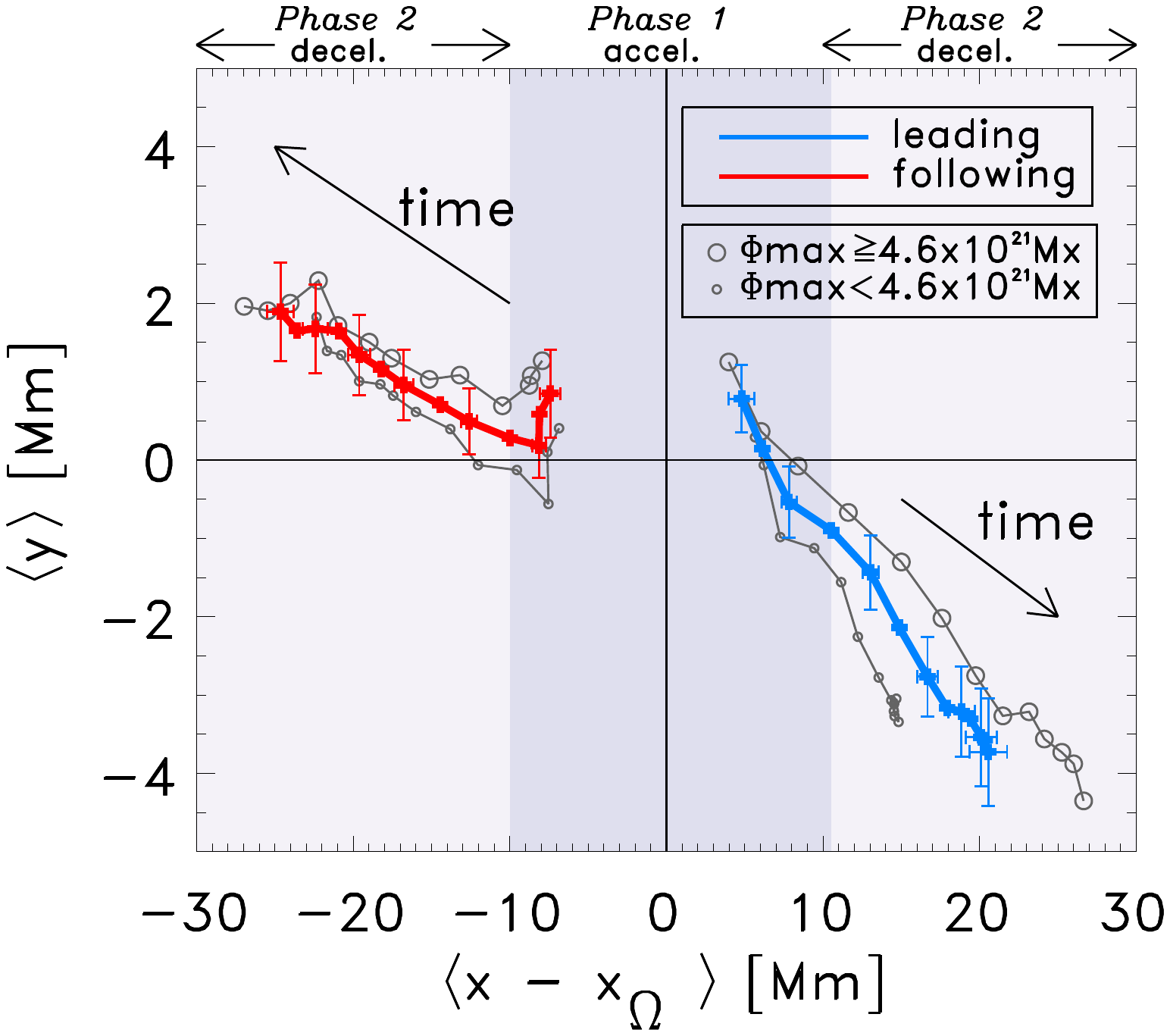}\\ 
\caption{
Average over the position of 153  positive (red) and negative (blue) polarities relative to the corrected centre of the map from $\tau=-18.4$~hrs (three time intervals, \texttt{TI-03}) before the emergence time, until $\tau=2.1$~days after (\texttt{TI+09}).
The centre of each of the maps were tracked at the Carrington rotation rate \citep{Snodgrass1984}.
We corrected the centre of the map by subtracting the displacement due to difference between the quiet-Sun plasma rotation rate $x_\Omega = R_\odot \Omega(\lambda) \cos(\lambda) \Delta \tau$, where $\lambda$ is the latitude of the centre of the Postel projected map (see Table~A.1 in \cite{Schunkeretal2016}).
The  blue and red curves cover the time intervals from \texttt{TI-03} to \texttt{TI+09}.
The grey lines with large (small) circles shows the motion of the polarities belonging to regions with maximum flux higher (lower) than the median flux.
The leading polarity moves south on average, but in the last few time intervals shown here, the low-flux polarities begin to move north, overlapping their previous path.
The shaded regions indicate Phase~1, when the separation speed between the polarities increases, and Phase~2 when the separation speed decreases (see Fig.~\ref{fig:vxvytime} and Sect.~\ref{sect:phases}).
}
\label{fig:bpposmap}
\end{figure*}

Figure~\ref{fig:bpposmap} shows the average motion of the leading $(x_\ell(\tau),y_\ell(\tau))$ and following $(x_f(\tau),y_f(\tau))$ polarity from about twelve hours prior to emergence and up to two days after emergence. 
At each time interval, we measure the position of the polarities in each EAR and then average these positions over all EARs.

\cite{Schunkeretal2016} showed that the east-west separation in the first day was antisymmetric about the latitudinal differential rotation rate at the surface of the Sun.
For each active region we subtracted the displacement in the  $x$-direction (east-west) due to the difference between the local plasma rotation rate \citep{Snodgrass1984} and the Carrington rotation rate at which each region was tracked.   
Figure~\ref{fig:bpposmap} shows that the east-west motion of the polarities is antisymmetric about the centre of the emerging active region.  
As expected, we found that, on average, the leading polarity moves in the prograde direction (positive $x$-direction) and  towards the equator (negative $y$-direction), and the following polarity moves in the retrograde direction (negative $x$-direction) and towards the pole (positive $y$-direction).  

It is well established that the leading and following polarities of active regions tend to move apart in the early stages of evolution. 
Here, we showed  that the largest separation occurs in the east-west direction.
We found that on average the polarities begin roughly east-west aligned \citep[as also reported by][]{KosovichevStenflo2008} and then move apart in the north-south direction. On average, the leading polarity moves equatorward further than the following polarity moves poleward in the first two days.

\section{Separation between the magnetic polarities as a function of time and magnetic flux}\label{sect:sepflux}

If the rising flux tube theory is valid, the separation of the polarities  reflects the underlying geometry of the tube as shown in \cite{Chenetal2017}.

Traditionally,  the separation between active region polarities has been described directly relative to one another. In this paper we examine the motion in the east-west and north-south  components independently. 
The  separation in the $y$-direction, $\delta y (\tau) =  y_\mathrm{l} (\tau) - y_\mathrm{f} (\tau)$, is negative when the leading  polarity is closer to the equator, and positive when it is closer to the pole, than the following  polarity.
The separation in the $x$-direction, $\delta x (\tau) =  x_\mathrm{l} (\tau) - x_\mathrm{f} (\tau)$, is positive when the leading polarity is in the prograde direction relative to the following polarity.
The total separation is then $\delta (\tau) = \sqrt{\delta x^2 + \delta y^2}$, and  the tilt angle is positive when the leading polarity is closer to the equator than the following polarity.

\begin{figure*}
\includegraphics[width=0.9\textwidth]{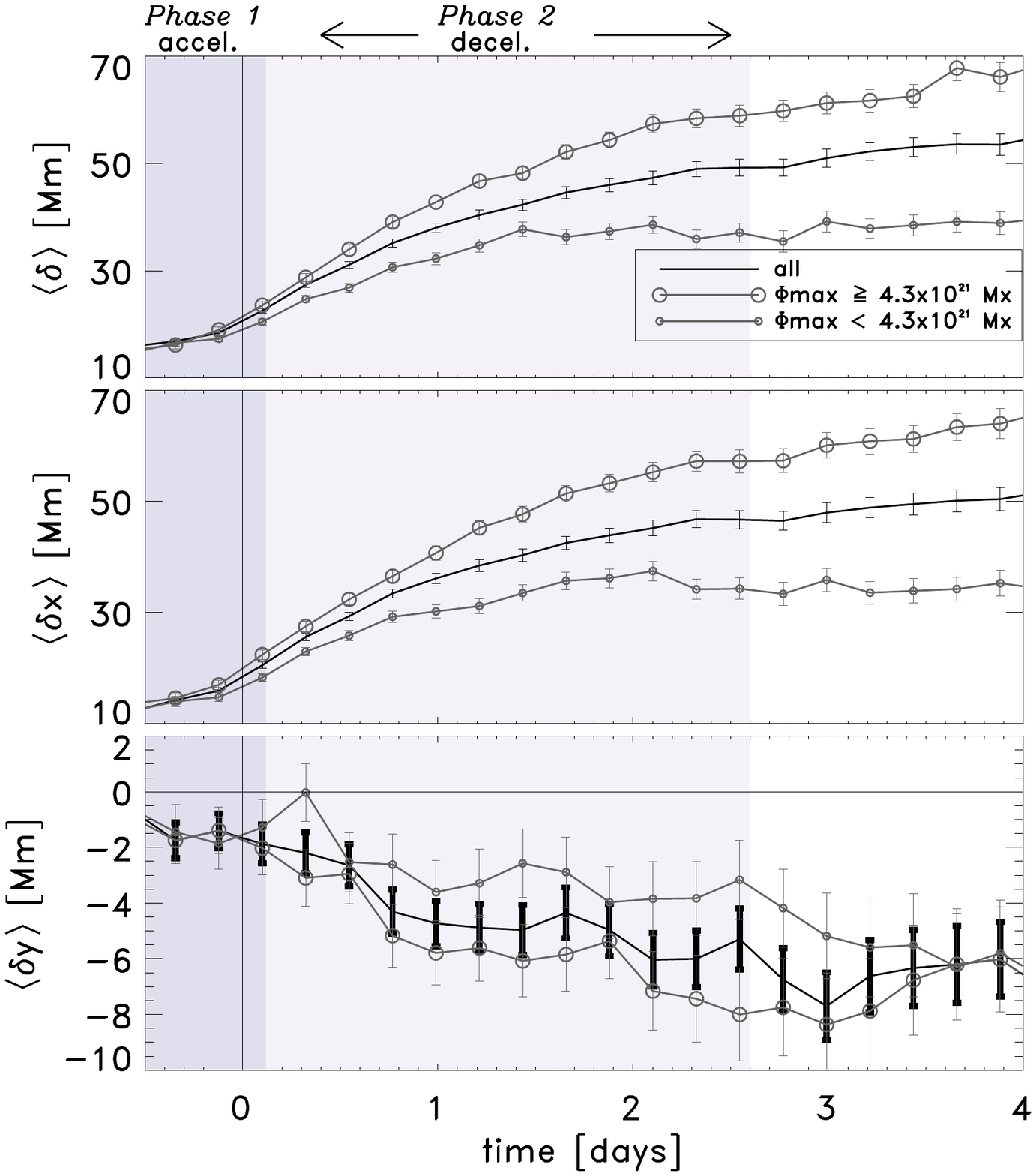} 
\caption{
Mean separation (top), $x$-separation (middle), and $y$-separation (bottom) as functions of time for all regions (black) and regions with a higher (lower) maximum flux than the median in large (small) grey circles.
The active regions are divided into high and low maximum flux by the median value, \medianflux.
Note that the uncertainties depend on  time.
Higher flux regions have a larger separation in the east-west direction at the time of emergence and this becomes more pronounced with time.
The shaded regions indicate two different phases of emergence, an increasing separation speed between the polarities, peaking at a value of \avesepspeed , followed by a decreasing separation speed (see Fig.~\ref{fig:vxvytime} and Sect.~\ref{sect:phases}).
 }
\label{fig:disttimemxflx}
\end{figure*}

The separation between the polarities is known to be related to the active region area. For example, \citet{WangSheeley1989} showed that the mean  separation distance between active region polarities  is a good proxy for the total magnetic flux in the region,  assuming active region flux is proportional to size in the intensity continuum. We found that a clearly flux-dependent component of separation occurs only in the east-west direction  (see Fig.~\ref{fig:disttimemxflx}) and that the separation increases until about two days after emergence. 
There is a systematic difference in the north-south separation, with higher flux regions being more separated than lower flux regions, but within the uncertainties it is not significant, and the lack of clear dependence on flux translates to the tilt angles.

\section{Separation speed between the magnetic polarities as a function of time and magnetic flux}\label{sect:vxvy}

The separation speed of the polarities in the east-west and north-south directions can help us to understand what forces are driving the separation. For example, the Coriolis force acting on east-west motions would result in an acceleration of the separation in the north-south direction, and a deceleration, would suggest magnetic tension effects are dominating in the north-south direction.

\begin{figure*}
\includegraphics[width=0.9\textwidth]{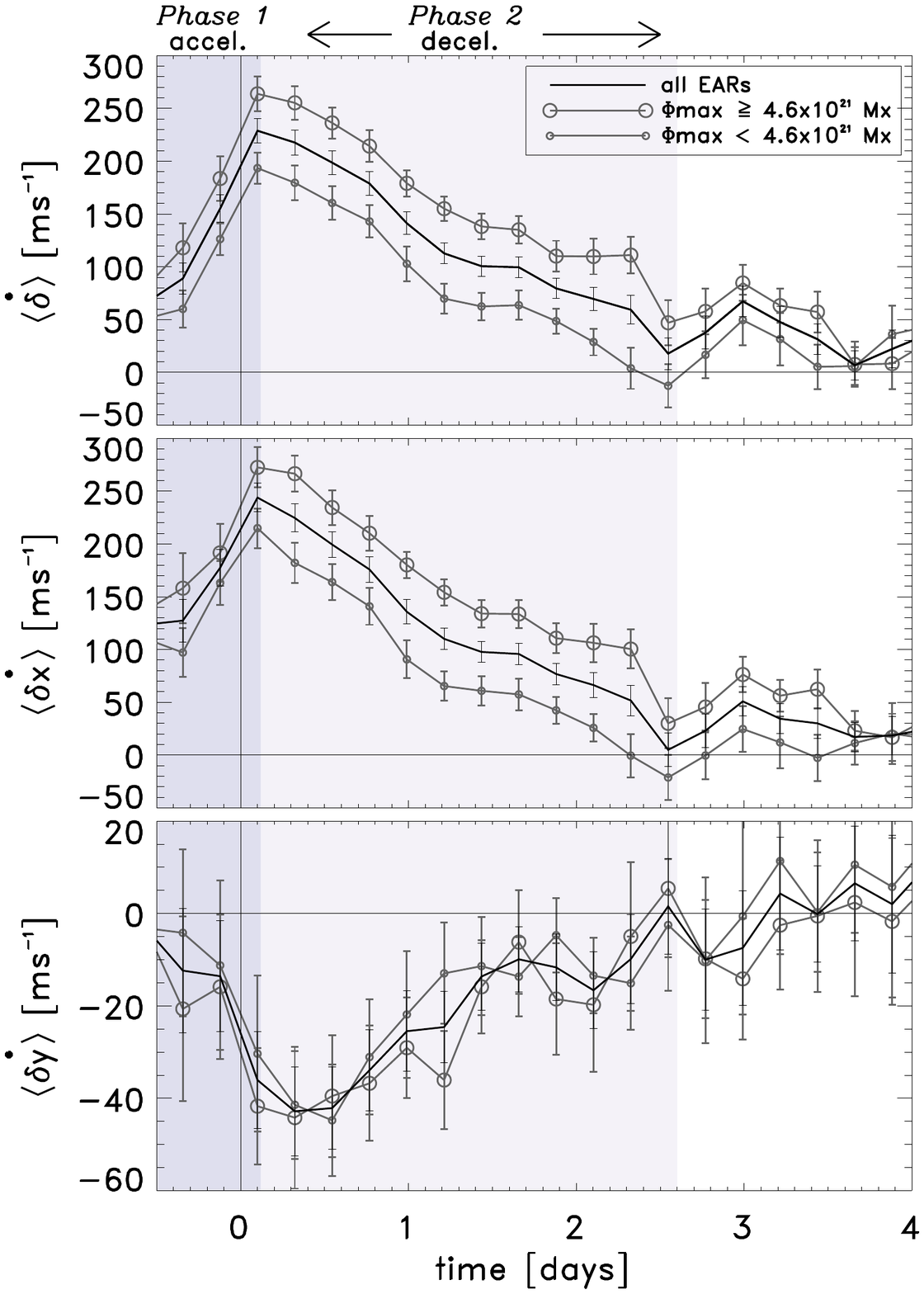} 
\caption{Mean separation speed (top), in the east-west direction (middle) and north-south direction (bottom).Higher flux regions have a larger east-west separation speed at the time of emergence than lower flux regions. There is no dependence of the gradient on flux.
The shaded regions indicate two different phases of emergence, where the speed of separation between the polarities is first increasing (accelerating) and then decreasing (decelerating) (see Fig.~\ref{fig:vxvytime} and Sect.~\ref{sect:phases}). }
\label{fig:vxvytime}
\end{figure*}

We estimate the separation speed numerically using 
$\dot{\delta} (\tau_i) = \left( \delta (\tau_{i+1}) - \delta (\tau_{i-1}) \right) / \left( \tau_{i+1} - \tau_{i-1} \right)$, where $i$ is the temporal index, and similarly for the 
north-south separation speed numerically  $\delta \dot{y}(\tau_i) = \left( \delta y(\tau_{i+1}) - \delta y(\tau_{i-1}) \right) / \left( \tau_{i+1} - \tau_{i-1} \right)$. The $\dot{}$ represents the time derivative. We use the analogous estimate for the east-west separation speed $\delta \dot{x}$.
Figure~\ref{fig:disttimemxflx} shows that the separation increases fastest in the first 1-2 days after emergence and is almost constant 4~days after emergence.

At each time interval, we first measured the separation speed of the polarities in each EAR and then average over the separation speed of all the EARs.
We then followed the same procedure for those with flux lower than or equal to the median, and then for those with flux higher than the median.

We measure  an average east-west separation speed of \sepspeedoneday  in the first day after emergence, which is faster compared to \citet{Schunkeretal2016} who measured a  separation speed of \sepspeed in the first day.
We emphasise that our technique to measure the separation speed is different to the method of \citet{Schunkeretal2016}, and suspect that the large difference in values is due to the simplified method used in \citet{Schunkeretal2016} where the authors only used a latitudinal average of the line-of-sight magnetic field at each time interval, and not the full map.
One example is AR~11310 presented in Fig.~10 of \citet{Schunkeretal2016}, where they measure an east-west separation speed of $168$~ms$^{-1}$, but with the measurement technique presented here we measure a separation speed of $319$~ms$^{-1}$. This is predominantly due to the trailing polarity expanding outside of the latitude range used for the averaging in the earlier paper.

Figure~\ref{fig:vxvytime} shows that the east-west and the north-south separation speed  peak at about 0.1~days after the emergence time, from whence the speeds decrease in magnitude at a rate independent of flux. 
This is consistent with an east-west oriented flux tube  rising through the surface:  the separation speed is fastest as the apex of the tube breaks the surface, the polarities then slow down to reach a null separation speed once they are above their anchoring depths  \citep[e.g.][]{Chenetal2017}. 

The east-west separation speed is dependent on flux, with higher flux region polarities separating faster in the east-west direction than lower flux regions. \cite{Birchetal2016} found that active regions emerge with rise speeds on the order of the convective velocity. 
Assuming that all flux tubes are rising at speeds comparable to convective velocities, then flux tubes with larger cross-sections, and thus larger surface area and flux, will appear to move apart faster than lower-flux tube with a smaller cross-sectional area.
Our finding that the east-west separation speed is dependent on flux is consistent with high-flux tubes having larger cross-sections than lower-flux tubes.
We found that after about 0.1~days the separation speed started to decrease and the flux dependence persisted. 

\subsection{The phases of emergence}\label{sect:phases}

\begin{figure*}
\includegraphics[width=0.9\textwidth]{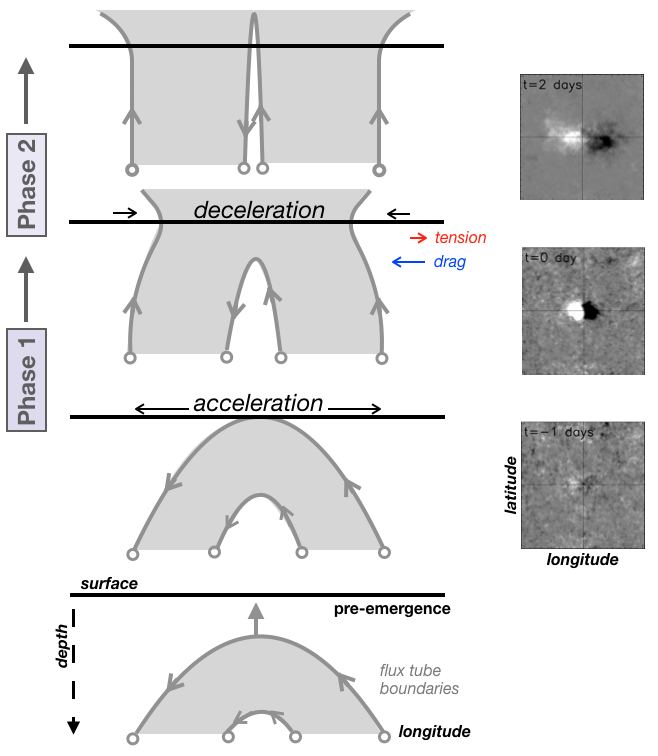} 
\caption{
Sketch of the emergence process of an anchored flux tube. Proceeding from the bottom panel upwards, a flux tube rises from some anchoring depth, $z_a$, towards the surface. Phase~1 begins about $\tau=-1$~days  when the opposite polarities become visible at the surface with an increasing separation speed (acceleration). 
We attribute this to the shape of the rising flux tube.
Phase~2 begins when the separation speed starts decreasing (deceleration) and ends when the polarities have stopped separating and lie above the sub-surface anchored foot-points. 
We attribute this to a combination of the magnetic tension force acting to straighten the magnetic field after it has expanded above the surface causing the polarities to separate, and the opposing drag force acting on the moving polarities.
 }
\label{fig:ecartoon}
\end{figure*}

Our results reveal two clear phases of emergence defined by the sign of the slope of the separation speed (acceleration or deceleration in Fig.~\ref{fig:vxvytime}) that distinguishes two clear phases of the emergence process:
\begin{itemize}
\item \textbf{Phase~1} is defined by the increasing speed of the separation  between the polarities. On average, active region polarities emerge east-west aligned marking the beginning of Phase~1, and lasting until about 0.1~days after the emergence time, when the separation speed ceases to increase. 
We interpret this as the apex of a rising east-west aligned flux tube breaking the surface.
\item \textbf{Phase~2} is defined by a decrease in separation speed, beginning about 0.1~days after emergence until the polarities stop separating about $2.5\--\,3$~days after the time of emergence. We propose that the footpoints of the tube are anchored, and the magnetic tension and the drag force acting on the magnetic field are the dominant forces \cite[as in][]{Chenetal2017} throughout Phase~2. 
The magnetic tension acts to straighten the magnetic field lines, after the flux tube has expanded above the surface, and the drag force \emph{opposes} the motion of the magnetic field.
\end{itemize}
We illustrate the phases in Fig.~\ref{fig:ecartoon}.
Given that active regions typically emerge with rise velocities on the order of the convective velocities \citep[$<70$ ms$^{-1}$, see ][]{Birchetal2016}, the flux dependence of the separation and separation speed may indicate that tubes with higher flux have a larger cross-section and radius of curvature. 
The acceleration and deceleration does not depend significantly on flux.
We do not observe any oscillations in the relative separation of the polarities (and as far as we know none have been reported in the literature), which would be expected if the dominant forces were the inertia and the magnetic tension.

\section{Scatter in the motion of the magnetic polarities}\label{sect:scatsepflux}

Models have shown that flux tubes with lower magnetic flux are more sensitive to near-surface flows as they rise \citep[e.g.][]{LongcopeFisher1996,Weberetal2011}, and therefore, the observed  motion of the lower flux polarities are predicted to have more scatter. 
This scatter is essential to producing a varying solar cycle amplitude in Babcock-Leighton and surface flux transport models  \citep[e.g.][]{KarakMiesch2017}.
We tested this theory by looking at the scatter of the separation between the polarities in high and low flux active regions. Figure~7 in \cite{Schunkeretal2016}  shows that  for lower maximum flux regions the flux stops increasing after about two days. Therefore, we can directly compare only the first two days after emergence between lower and higher flux regions. After this the lower flux regions have started to dissipate.

\begin{figure*}
\includegraphics[width=0.48\textwidth]{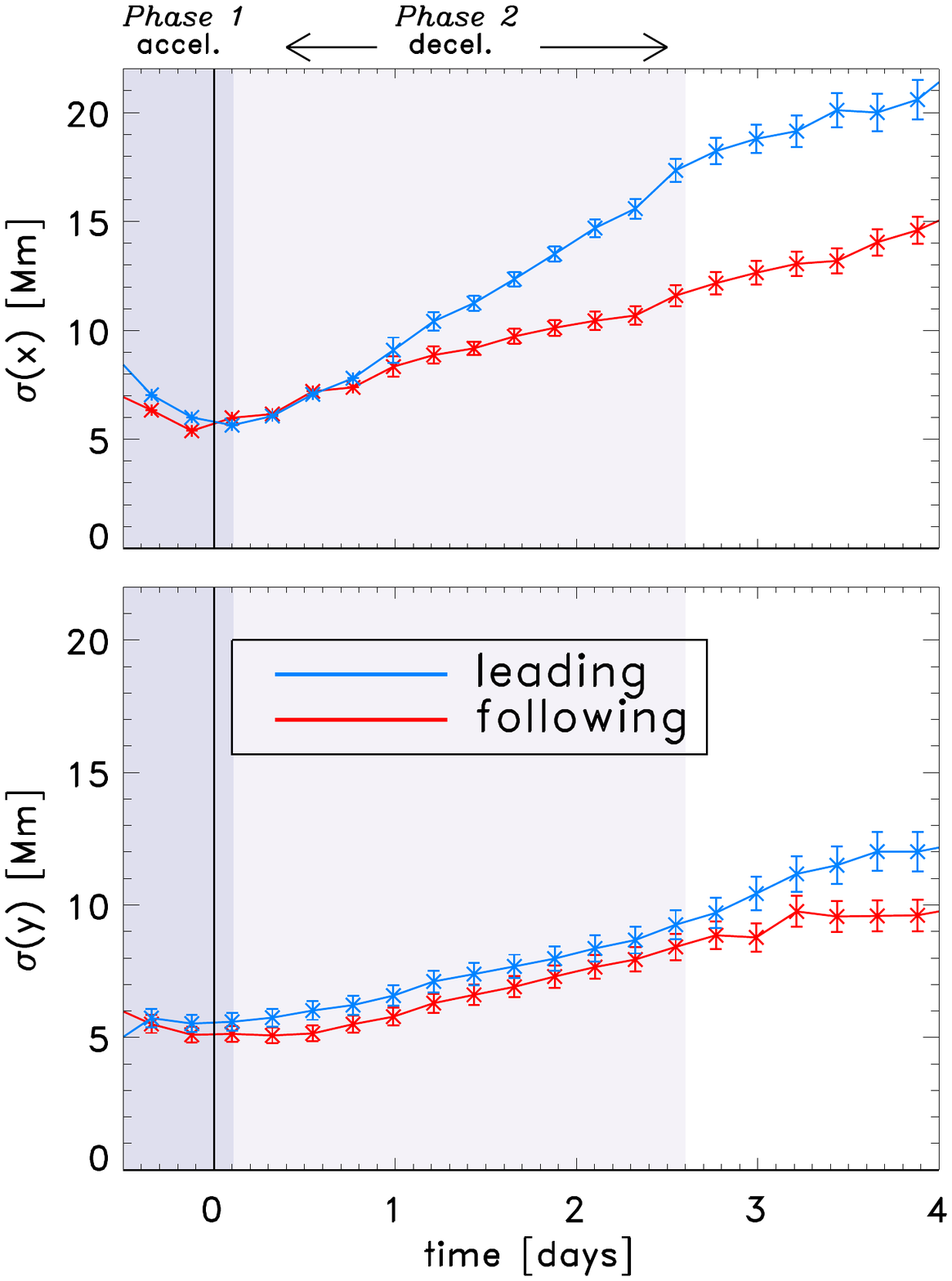}
\includegraphics[width=0.48\textwidth]{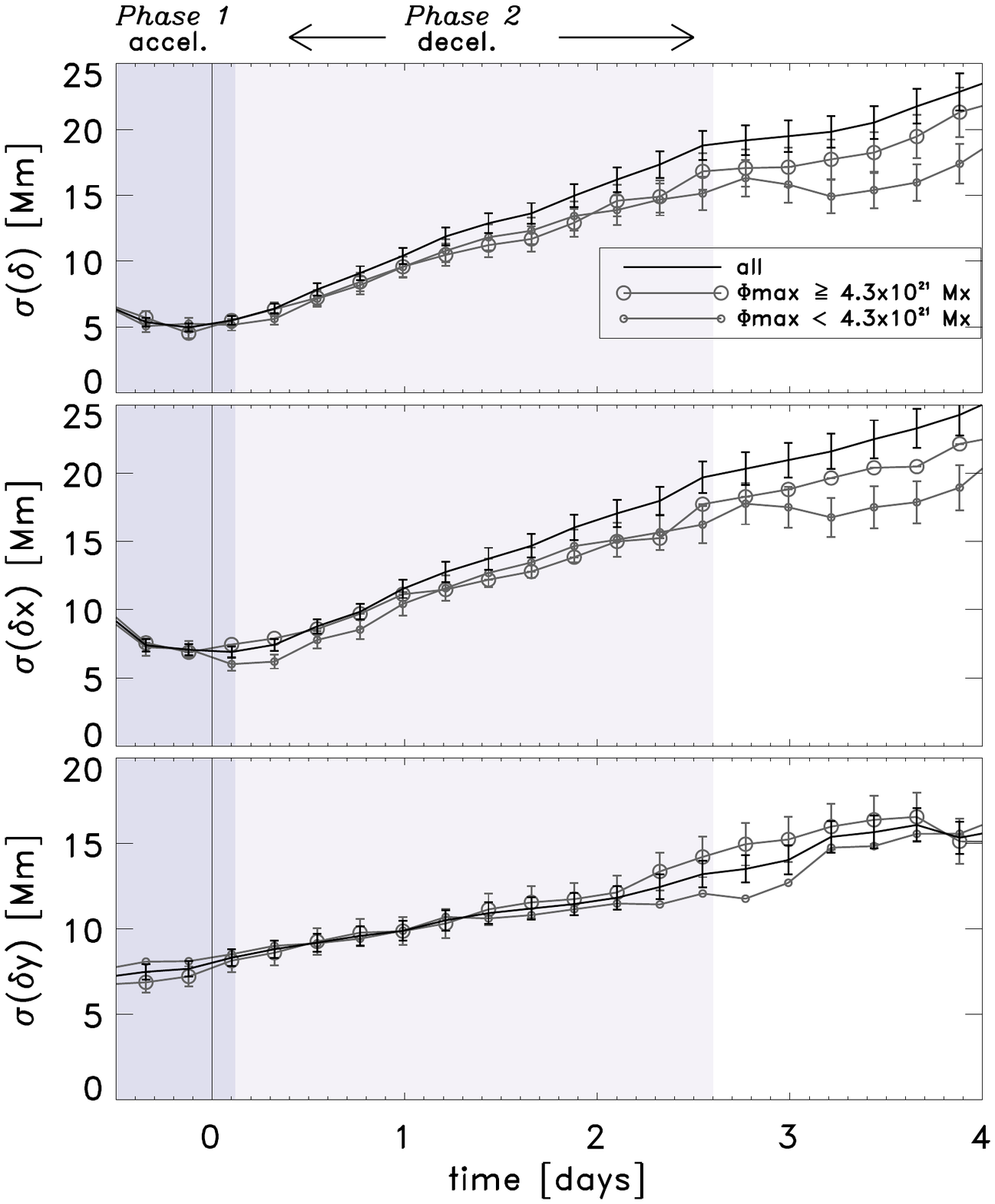} 
\caption{
Left: Standard deviation in the position of the leading (blue) and following (red) polarity position as a function of time. 
The error bars are the standard deviation of the sample standard deviation at each time interval as described by Eq.~\ref{eqn:sdscat} in Appendix~\ref{app:sd_scatter}.
To explain the difference in scatter, we suggest that the following polarity has less scatter because it is more diffuse than the leading polarity and the centre of gravity is not as affected by the buffeting from  supergranules. 
Right: Standard deviation of the separation, $x$-separation and $y$-separation  of the polarities (from top to bottom panels) as a function of time for all EARs (black), EARs with a higher (lower) maximum flux than the median in large (small) grey circles.
The EARs are divided into high and low maximum flux by the median value, \medianflux .
The standard deviation of the sample standard deviation at each time interval is given by Eq.~\ref{eqn:sdscat} in Appendix~\ref{app:sd_scatter}. 
The uncertainty in the separation of the polarities is largely independent of flux, and on the scale of supergranulation, which suggests that the scatter is not dependent on magnetic tension, but buffeting by supergranulation.
The shaded regions indicate two different phases of emergence, an increasing separation speed between the polarities followed by a decreasing separation speed, defined in Fig.~\ref{fig:vxvytime} and Sect.~\ref{sect:phases}.
}
\label{fig:bpscat}
\end{figure*}

Figure~\ref{fig:bpscat} (left) shows the standard deviation in the positions of the leading and following polarities in Fig.~\ref{fig:bpposmap}.  
The scatter is systematically larger for the leading polarity than for the following polarity. 
This may be due to the statistically significant flux dependence of the position of the leading polarity, which is not evident for the following polarity.
It may also be due to the fact that the following polarity becomes more diffuse in time \cite[e.g.][ and references therein]{vDGGreen2015}, and so small changes in flux distribution will not significantly change the location of the local maximum.

Figure~\ref{fig:bpscat} (right) shows that the standard deviation in the total separation of both high and low flux regions increases linearly, and are similar up to two days after emergence. 
There is a systematic but insignificant difference between the scatter in the separation of the high and low flux regions, mostly from the standard deviation in the east-west separation. 
The standard deviation in separation in the north-south direction also increases linearly, and is independent of the maximum flux of the region.
This is consistent with advection by the random background motions operating on  scales smaller than the distribution of the flux.

The  standard deviation shown for the polarity positions and separation between then is on the order of tens of megametres, similar to supergranulation spatial scales \citep[20-40~Mm, see ][]{Langfellneretal2015_Vorticity}, and the increase in the standard deviation as a function of time suggests a random walk.
This supports a model of emergence where supergranulation plays a key role in buffeting the polarities, regardless of flux strength.
We found that the scatter in separation is independent of flux. This is in contrast to the thin flux-tube simulations of \cite[e.g.][]{LongcopeFisher1996,Weberetal2011}. Although this theory is only valid deeper than $\approx 20$~Mm below the surface, these results show that the near surface convection plays a significant role in emergence.

\section{Summary}\label{sect:summary}

We have identified two clear phases of emergence:
\begin{itemize}
\item Phase~1:  the separation between the polarities is accelerating. This phase begins when the surface polarities are first detectable and lasts until about 0.1 days after emergence, and 
\item Phase~2: the separation between the polarities is decelerating. This phase begins when Phase~1 ends and lasts until about 2.5~days after emergence, when separation speed is close to zero.
\end{itemize}

Along with  our finding that the polarities of higher-flux tubes with a larger cross-section separate faster than lower-flux tubes with a smaller cross-section, Phase~1 is consistent with the apex of a magnetic flux tube breaking the surface. 
Phase~2 is consistent with magnetic tension straightening the field lines so that the surface polarities lie directly above the anchored footpoints below the surface, as in \citep[][]{Chenetal2017} countered by the opposing drag force. 

During Phase 2, the increase in the scatter in the location of the polarities is on supergranulation length scales (tens of Mm) and time scales (one to two days).
We also found a lack of significant flux dependence, that suggests that the scatter is due to buffeting by supergranulation.
\begin{acknowledgements}
The German Data Center for SDO, funded by the German Aerospace Center (DLR), provided the IT infrastructure for this project.  Observations courtesy of NASA/SDO and the HMI science teams. This work utilized the Pegasus workflow management system.
DCB is supported by the Solar Terrestrial program of the US National Science Foundation (grant
AGS-1623844) and by the Heliophysics Division of NASA (grants 80NSSC18K0068 and 80NSSC18K0066).
\end{acknowledgements}

\bibliographystyle{aa} 
\bibliography{ears.bib} 

\appendix

\clearpage 
\onecolumn
\section{Additional emerging solar active regions added to the SDO/HEAR survey}\label{app1}

\begin{longtable}{ l c c c c c | c c | c c }
\caption{Emerging active region and control region tracking locations and emergence time.} \label{tab:ears} \\
 \, \, AR & emergence time &  lat.  & lon. & CMD  & $P$ & CR emergence time & CR lon.  &  $\Delta B0$ &  $\Delta T$ \\
 \, \, \, \#  & [TAI] & [$^\circ$] & [$^\circ$] & [$^\circ$] &  & [TAI] & [$^\circ$] & [$^\circ$] & [days] \\
\hline
\endfirsthead
\multicolumn{3}{l}{{\bfseries \tablename\ \thetable{} -- continued from previous page}} \\
 \, \,  AR & emergence time &  lat.  & lon. & CMD  & $P$ & CR emergence time & CR lon.  &  $\Delta B0$ &  $\Delta T$ \\
 \, \, \, \#  &[TAI] & [$^\circ$] & [$^\circ$] & [$^\circ$] &  & [TAI] & [$^\circ$] & [$^\circ$] & [days] \\
\hline
\endhead
11626   &  2012.12.03\_01:36:00   &     12.5   &    299.0   &    -49.2   &  3   &  2012.12.05\_01:36:00   &    272.6   &     -0.3   &    2.0 \\ 
11627   &  2012.12.03\_20:00:00   &    -15.2   &    319.7   &    -18.5   &  1   &  2012.12.07\_19:59:15   &    267.0   &     -0.5   &    4.0 \\ 
11631*   &  2012.12.12\_02:36:00   &     19.5   &    222.9   &     -6.2   &  2   &  2012.12.08\_02:36:00   &    275.6   &      0.5   &   -4.0 \\ 
11640*   &  2012.12.29\_15:24:00   &     27.8   &    319.3   &    -38.9   &  0   &  2012.12.31\_15:24:00   &    292.9   &     -0.2   &    2.0 \\ 
11645*   &  2013.01.02\_20:12:00   &    -13.3   &    290.4   &    -12.4   &  0   &  2012.12.29\_20:12:00   &    343.1   &      0.5   &   -4.0 \\ 
11670*   &  2013.02.06\_22:36:00   &     18.4   &    159.5   &    -41.1   &  2   &  2013.02.25\_00:00:00   &    281.7   &     -0.7   &   18.1 \\ 
11675*   &  2013.02.16\_06:36:00   &     12.5   &     34.2   &    -43.5   &  0   &  2013.02.18\_06:36:45   &      7.9   &     -0.1   &    2.0 \\ 
11680*   &  2013.02.24\_16:48:00   &    -28.7   &    273.6   &    -53.2   &  0   &  2013.02.26\_16:48:45   &    247.2   &     -0.0   &    2.0 \\ 
11686*   &  2013.03.02\_08:12:00   &    -13.2   &    262.0   &      9.6   &  2   &  2013.03.04\_08:12:45   &    235.7   &     -0.0   &    2.0 \\ 
11696*   &  2013.03.11\_10:24:00   &      4.4   &     90.5   &    317.8   &  1   &  2013.03.20\_12:00:00   &    331.0   &      0.2   &    9.1 \\ 
11697   &  2013.03.13\_13:00:00   &     14.7   &    107.7   &      2.8   &  1   &  2013.03.22\_12:00:00   &    349.6   &      0.2   &    9.0 \\ 
11699*   &  2013.03.17\_00:24:00   &    -15.8   &     91.4   &     32.3   &  0   &  2013.03.05\_12:00:00   &    243.2   &     -0.1   &  -11.5 \\ 
11702*   &  2013.03.21\_02:12:00   &      8.3   &     14.9   &      9.5   &  0   &  2013.03.23\_02:12:00   &    348.5   &      0.1   &    2.0 \\ 
11703*   &  2013.03.21\_11:48:00   &    -23.8   &     39.5   &     39.3   &  2   &  2013.03.23\_11:48:00   &     13.1   &      0.1   &    2.0 \\ 
11706   &  2013.03.27\_01:24:00   &     -6.5   &    268.7   &    -18.0   &  1   &  2013.04.03\_01:23:15   &    176.4   &      0.4   &    7.0 \\ 
11707   &  2013.03.28\_11:48:00   &    -10.7   &    229.0   &    -38.8   &  0   &  2013.03.26\_11:48:00   &    255.4   &     -0.1   &   -2.0 \\ 
11712   &  2013.03.30\_19:00:00   &      1.6   &    183.0   &    -54.5   &  2   &  2013.04.01\_18:59:15   &    156.6   &      0.1   &    2.0 \\ 
11718*   &  2013.04.05\_15:24:00   &     22.0   &    109.6   &    -50.6   &  0   &  2013.04.03\_15:24:00   &    136.0   &     -0.1   &   -2.0 \\ 
11726*   &  2013.04.19\_09:48:00   &     12.6   &    322.1   &    -16.4   &  2   &  2013.04.17\_09:48:45   &    348.5   &     -0.2   &   -2.0 \\ 
11736   &  2013.04.30\_19:00:00   &     -7.1   &    135.1   &    -53.0   &  2   &  2013.05.02\_18:59:15   &    108.7   &      0.2   &    2.0 \\ 
11750*   &  2013.05.15\_01:48:00   &    -10.3   &    359.8   &      0.5   &  3   &  2013.05.24\_01:48:45   &    240.7   &      1.0   &    9.0 \\ 
11752   &  2013.05.15\_17:36:00   &     18.7   &      1.7   &     11.1   &  2   &  2013.05.30\_12:00:00   &    166.3   &      1.8   &   14.8 \\ 
11764*   &  2013.06.02\_01:24:00   &     12.2   &    128.4   &      7.0   &  3   &  2013.06.05\_12:00:00   &     82.9   &      0.4   &    3.4 \\ 
11776*   &  2013.06.18\_12:24:00   &     11.7   &    252.1   &    -11.5   &  1   &  2013.06.16\_12:24:45   &    278.5   &     -0.2   &   -2.0 \\ 
11780*   &  2013.06.26\_11:00:00   &     -8.3   &    140.3   &    -18.1   &  2   &  2013.06.28\_10:59:15   &    113.8   &      0.2   &    2.0 \\ 
11781*   &  2013.06.27\_23:48:00   &     22.3   &    128.3   &     -9.8   &  1   &  2013.07.06\_23:47:15   &      9.2   &      1.0   &    9.0 \\ 
11784*   &  2013.07.01\_11:24:00   &    -14.8   &     52.7   &    -39.3   &  3   &  2013.07.03\_11:24:45   &     26.2   &      0.2   &    2.0 \\ 
11786   &  2013.07.02\_00:00:00   &    -32.1   &     53.7   &    -31.4   &  0   &  2013.07.04\_00:00:00   &     27.2   &      0.2   &    2.0 \\ 
11789   &  2013.07.06\_13:36:00   &    -26.1   &    342.4   &    -42.2   &  2   &  2013.07.08\_13:36:45   &    315.9   &      0.2   &    2.0 \\ 
11802*   &  2013.07.24\_12:12:00   &     13.2   &    202.3   &     55.1   &  3   &  2013.07.17\_12:12:00   &    295.0   &     -0.6   &   -7.0 \\ 
11807*   &  2013.07.28\_10:36:00   &     28.9   &     91.6   &     -3.6   &  0   &  2013.07.13\_10:36:00   &    290.1   &     -1.3   &  -15.0 \\ 
11811   &  2013.07.31\_06:24:00   &      5.2   &      7.1   &    -50.7   &  1   &  2013.08.08\_12:00:00   &    258.2   &      0.5   &    8.2 \\ 
11813*   &  2013.08.06\_20:00:00   &    -13.1   &    320.7   &    -10.2   &  0   &  2013.08.11\_01:20:15   &    264.9   &      0.3   &    4.2 \\ 
11821   &  2013.08.14\_06:24:00   &      1.3   &    245.4   &     12.7   &  1   &  2013.08.10\_17:20:15   &    292.2   &     -0.2   &   -3.5 \\ 
11824*   &  2013.08.17\_07:36:00   &    -14.8   &    194.8   &      2.4   &  1   &  2013.08.26\_12:00:00   &     73.4   &      0.3   &    9.2 \\ 
11829*   &  2013.08.20\_17:00:00   &      4.2   &    190.0   &     42.4   &  3   &  2013.08.23\_17:00:00   &    150.3   &      0.1   &    3.0 \\ 
11831*   &  2013.08.21\_06:48:00   &     13.5   &    165.2   &     25.2   &  2   &  2013.08.24\_06:47:15   &    125.5   &      0.1   &    3.0 \\ 
11833   &  2013.08.22\_08:48:00   &     19.8   &     96.9   &    -28.7   &  4   &  2013.08.26\_12:00:00   &     42.3   &      0.1   &    4.1 \\ 
11842   &  2013.09.11\_04:36:00   &      4.9   &    259.9   &     36.1   &  1   &  2013.09.07\_04:35:15   &    312.7   &      0.0   &   -4.0 \\ 
11843   &  2013.09.17\_08:00:00   &      0.8   &    127.4   &    -15.3   &  0   &  2013.09.21\_08:00:00   &     74.6   &     -0.1   &    4.0 \\ 
11849*   &  2013.09.19\_13:00:00   &     20.9   &     75.3   &    -38.2   &  1   &  2013.09.16\_12:00:00   &    115.5   &      0.1   &   -3.0 \\ 
11855*   &  2013.09.30\_01:00:00   &    -11.5   &    305.8   &    -29.1   &  2   &  2013.10.07\_01:00:00   &    213.4   &     -0.4   &    7.0 \\ 
11867*   &  2013.10.09\_05:00:00   &     23.2   &    180.3   &    -33.7   &  0   &  2013.10.25\_05:00:45   &    329.2   &     -1.2   &   16.0 \\ 
11874*   &  2013.10.17\_04:00:00   &    -10.8   &     76.0   &    -33.0   &  1   &  2013.10.25\_15:00:00   &    324.5   &     -0.7   &    8.5 \\ 
11878   &  2013.10.19\_15:24:00   &     -9.9   &    110.1   &     33.7   &  3   &  2013.10.24\_12:00:00   &     46.0   &     -0.4   &    4.9 \\ 
11886   &  2013.10.28\_05:00:00   &     14.9   &    307.4   &    -15.9   &  3   &  2013.10.01\_05:00:45   &    303.6   &      2.0   &  -27.0 \\ 
11894   &  2013.11.07\_08:48:00   &     -7.0   &    200.3   &     10.8   &  3   &  2013.11.11\_12:00:00   &    145.8   &     -0.5   &    4.1 \\ 
11902   &  2013.11.14\_13:36:00   &     19.7   &     81.2   &    -13.3   &  3   &  2013.11.19\_13:35:15   &     15.3   &     -0.6   &    5.0 \\ 
11910*   &  2013.11.27\_13:12:00   &      1.5   &    276.3   &     -7.1   &  1   &  2013.11.25\_13:11:15   &    302.7   &      0.3   &   -2.0 \\ 
11911   &  2013.11.30\_01:12:00   &    -11.8   &    220.8   &    -29.7   &  2   &  2013.11.28\_01:12:45   &    247.1   &      0.3   &   -2.0 \\ 
11915*   &  2013.12.03\_05:48:00   &    -29.6   &    206.9   &     -1.5   &  2   &  2013.11.26\_00:00:00   &    302.3   &      0.9   &   -7.2 \\ 
11922*   &  2013.12.10\_02:00:00   &     10.4   &    122.9   &      4.7   &  4   &  2013.12.23\_12:00:00   &    306.2   &     -1.7   &   13.4 \\ 
11924*   &  2013.12.10\_03:12:00   &    -13.2   &    152.0   &     34.5   &  2   &  2013.12.18\_12:00:00   &     41.8   &     -1.1   &    8.4 \\ 
11932   &  2013.12.18\_18:00:00   &      3.7   &    328.6   &    -35.4   &  3   &  2013.12.22\_17:59:15   &    275.9   &     -0.5   &    4.0 \\ 
11945   &  2014.01.02\_06:12:00   &     11.3   &    144.0   &    -28.9   &  1   &  2013.12.22\_06:11:15   &    288.9   &      1.3   &  -11.0 \\ 
11946*   &  2014.01.04\_10:36:00   &      9.8   &     99.9   &    -44.3   &  3   &  2013.12.26\_17:20:15   &    214.7   &      1.0   &   -8.7 \\ 
11951   &  2014.01.09\_09:48:00   &    -12.8   &     44.5   &    -34.3   &  1   &  2014.01.01\_12:00:00   &    148.7   &      0.9   &   -7.9 \\ 
11962   &  2014.01.19\_07:48:00   &    -37.2   &    279.6   &    -28.6   &  0   &  2014.01.21\_07:47:15   &    253.3   &     -0.2   &    2.0 \\ 
11969*   &  2014.01.30\_19:24:00   &    -10.5   &    159.8   &      2.8   &  1   &  2014.01.17\_12:00:00   &    335.1   &      1.1   &  -13.3 \\ 
11978*   &  2014.02.10\_07:24:00   &      5.6   &     34.0   &     15.3   &  1   &  2014.01.31\_07:24:00   &    165.7   &      0.6   &  -10.0 \\ 
11988*   &  2014.02.21\_23:00:00   &    -10.4   &    175.4   &    -50.0   &  1   &  2014.02.16\_12:00:00   &    247.3   &      0.2   &   -5.5 \\ 
11992   &  2014.02.25\_20:36:00   &    -20.2   &    137.1   &    -36.8   &  3   &  2014.02.23\_20:35:15   &    163.5   &      0.0   &   -2.0 \\ 
12003*   &  2014.03.09\_17:00:00   &      5.9   &     11.4   &     -6.5   &  2   &  2014.02.20\_12:00:00   &    238.1   &      0.2   &  -17.2 \\ 
12011*   &  2014.03.18\_08:24:00   &     -7.0   &    276.4   &     12.4   &  1   &  2014.03.30\_12:00:00   &    116.2   &      0.5   &   12.2 \\ 
12029*   &  2014.04.01\_08:12:00   &     17.8   &     26.9   &    -52.5   &  1   &  2014.04.07\_08:12:45   &    307.8   &      0.4   &    6.0 \\ 
12039   &  2014.04.15\_15:12:00   &     23.9   &    234.8   &    -16.0   &  1   &  2014.04.18\_15:12:45   &    195.2   &      0.2   &    3.0 \\ 
12041   &  2014.04.15\_15:36:00   &    -20.7   &    262.3   &     11.7   &  0   &  2014.04.13\_12:00:00   &    290.7   &     -0.2   &   -2.2 \\ 
12048*   &  2014.04.26\_12:12:00   &     19.5   &    124.4   &     17.2   &  3   &  2014.04.13\_12:12:00   &    296.1   &     -1.1   &  -13.0 \\ 
12062*   &  2014.05.10\_06:24:00   &     -6.4   &    290.5   &      5.2   &  4   &  2014.05.18\_12:00:00   &    181.6   &      0.9   &    8.2 \\ 
12064   &  2014.05.13\_23:12:00   &      8.3   &    194.6   &    -41.8   &  2   &  2014.05.21\_23:11:15   &     88.8   &      0.9   &    8.0 \\ 
12078   &  2014.05.31\_00:48:00   &    -18.4   &    327.4   &    -43.2   &  1   &  2014.05.28\_12:00:00   &      1.0   &     -0.3   &   -2.5 \\ 
12089*   &  2014.06.10\_21:36:00   &     17.6   &    195.9   &    -30.9   &  3   &  2014.05.20\_12:00:00   &    119.1   &     -2.5   &  -21.4 \\ 
12098   &  2014.06.23\_16:36:00   &     -8.3   &     21.0   &    -36.5   &  2   &  2014.06.25\_16:36:00   &    354.6   &      0.2   &    2.0 \\ 
12099   &  2014.06.26\_12:24:00   &    -17.5   &      5.2   &    -14.9   &  2   &  2014.06.24\_12:24:45   &     31.6   &     -0.2   &   -2.0 \\ 
12105   &  2014.06.28\_23:24:00   &     -7.1   &    307.8   &    -39.7   &  2   &  2014.06.26\_23:24:00   &    334.3   &     -0.2   &   -2.0 \\ 
12118   &  2014.07.17\_17:24:00   &      7.0   &    113.3   &     13.9   &  0   &  2014.07.16\_12:00:00   &    129.5   &     -0.1   &   -1.2 \\ 
12119*   &  2014.07.18\_11:12:00   &    -22.1   &     66.8   &    -22.8   &  1   &  2014.07.22\_11:12:00   &     13.9   &      0.4   &    4.0 \\ 
\end{longtable}

\tablefoot{
The left panel of the table lists the NOAA active region number,   emergence time, Carrington latitude, Carrington longitude, central meridian distance (CMD) at the time of emergence, and the $P$-factor. The middle panel lists the emergence time and  Carrington longitude of the control region. 
The right panel lists the difference in solar $B$-angle between the emergence time of the EAR and the CR, $\Delta B = \texttt{CRLT\_OBS}\mathrm{(CR)} - \texttt{CRLT\_OBS}\mathrm{(EAR)}$ where  \texttt{CRLT\_OBS} is the SDO/HMI keyword specifying the latitude at the centre of the map at the time of observation. 
The difference of the emergence time of the EAR and the CR rounded  to the nearest day is given by $\Delta T=\texttt{T\_REC}(\mathrm{CR}) - \texttt{T\_REC}(\mathrm{EAR})$, where \texttt{T\_REC} is the SDO/HMI keyword specifying the time the observation was made.
The star ($^*$) indicates  regions with a maximum flux larger than the median  of the first set published in \citet[\medianfluxsetone in ][]{Schunkeretal2016}. The median maximum flux of the active regions in the entire SDO/HEARS is $5.3\times 10^{21}$~Mx.
}

\clearpage 
\section{Time intervals}\label{app3}
\begin{table}
\caption{Mid-time of the time interval relative to the emergence time of the active region. Each time interval is 6.8~hours long.}
\begin{tabular}{l | c | c }
  time  & time & time \\
  interval  & [days] & [hours] \\
\hline
\texttt{TI-05} & $ -1.01$ & $-24.28$ \\
\texttt{TI-04} & $ -0.79$ & $-18.94$ \\
\texttt{TI-03} & $ -0.57$ & $-13.61$ \\
\texttt{TI-02} & $ -0.34$ & $ -8.27$ \\
\texttt{TI-01} & $ -0.12$ & $ -2.93$ \\
\texttt{TI+00} & $ 0.10$ & $ 2.41$ \\
\texttt{TI+01} & $ 0.32$ & $ 7.74$ \\
\texttt{TI+02} & $ 0.55$ & $13.08$ \\
\texttt{TI+03} & $ 0.77$ & $18.42$ \\
\texttt{TI+04} & $ 0.99$ & $23.76$ \\
\texttt{TI+05} & $ 1.21$ & $29.09$ \\
\texttt{TI+06} & $ 1.43$ & $34.43$ \\
\texttt{TI+07} & $ 1.66$ & $39.77$ \\
\texttt{TI+08} & $ 1.88$ & $45.11$ \\
\texttt{TI+09} & $ 2.10$ & $50.44$ \\
\texttt{TI+10} & $ 2.32$ & $55.78$ \\
\texttt{TI+11} & $ 2.55$ & $61.12$ \\
\texttt{TI+12} & $ 2.77$ & $66.46$ \\
\texttt{TI+13} & $ 2.99$ & $71.79$ \\
\texttt{TI+14} & $ 3.21$ & $77.13$ \\
\texttt{TI+15} & $ 3.44$ & $82.47$ \\
\texttt{TI+16} & $ 3.66$ & $87.81$ \\
\texttt{TI+17} & $ 3.88$ & $93.14$ \\
\texttt{TI+18} & $ 4.10$ & $98.48$ \\
\end{tabular}

\label{tab:ti}
\end{table}

\twocolumn
\section{Emerging solar active regions excluded from this analysis}\label{app2}
The anti-Hale active regions  and unreliable position measurements were identified by  visual inspection of the magnetograms.
We excluded 29 active regions from the analysis.\\

Anti-Hale's law regions: 
11194 11291 11326 11331 11574 12099  \\

Unreliable position measurement: 11074 11081 11154 11159 11223 11297 11322 11331 11381 11406 11414 11466 11531 11565 11703 11726 11776 11786 11802 11829 11842 11910 11924 12062

\clearpage
\section{Comparison of methods to identify the location of magnetic polarities}\label{app:methcomp}
\begin{table}
\caption{Positions of the centroids identified by the feature identification code for the leading $(x_l,y_l)$ and following $(x_f,y_f)$ polarities in the AR~11075 line-of-sight magnetic field map for different time averages of the 45~s datacube for \texttt{TI+02}. The standard deviation of the positions are shown in the bottom row. See also Fig.~\ref{fig:expos}.}
\begin{tabular}{l || c | c | c | c  }
    & $x_f$ & $y_f$ & $x_l$ & $y_l$ \\ 
    & [pix] & [pix] & [pix] & [pix] \\ 
\hline
 ave0   & 244.9611 & 260.2111 & 259.4323 & 251.7186 \\ 
 ave1   & 244.9684 & 260.2115 & 259.4247 & 251.7187 \\ 
 ave2   & 244.9700 & 260.2128 & 259.4258 & 251.7209 \\ 
 ave3   & 244.9610 & 260.2168 & 259.4218 & 251.7084 \\ 
 \hline  
 $\sigma$ &    0.005 &    0.003 &    0.004 &    0.006 \\ 
\end{tabular}

\label{tab:ar11075}
\end{table}

We made an estimate of the uncertainty in our method by  applying it to four different time averages of the 45~s cadence line-of-sight magnetic field maps in \texttt{TI+02} of AR~11075. Each time-averaged magnetic field map consists of averaging every fourth map, i.e. with a 3~minute cadence. The first average began  with the first frame, the second average began with the second frame, and so on. Table~\ref{tab:ar11075} shows the  measured position of the positive and negative polarity in each of these averages, in the $x$ and $y$-direction relative to the lower left corner pixel $(x,y)=(0,0)$,  by using the feature identification method. They differ by less than a percent of the pixel size, which is less than the typical change in position of the polarity from one time interval to the next, which is on the order of a pixel. Based on this experiment, we expect that the physical evolution of the emerging active  regions will be the dominant source of uncertainty, rather than any noise in the magnetograms.

\begin{figure}
\includegraphics[width=0.5\textwidth]{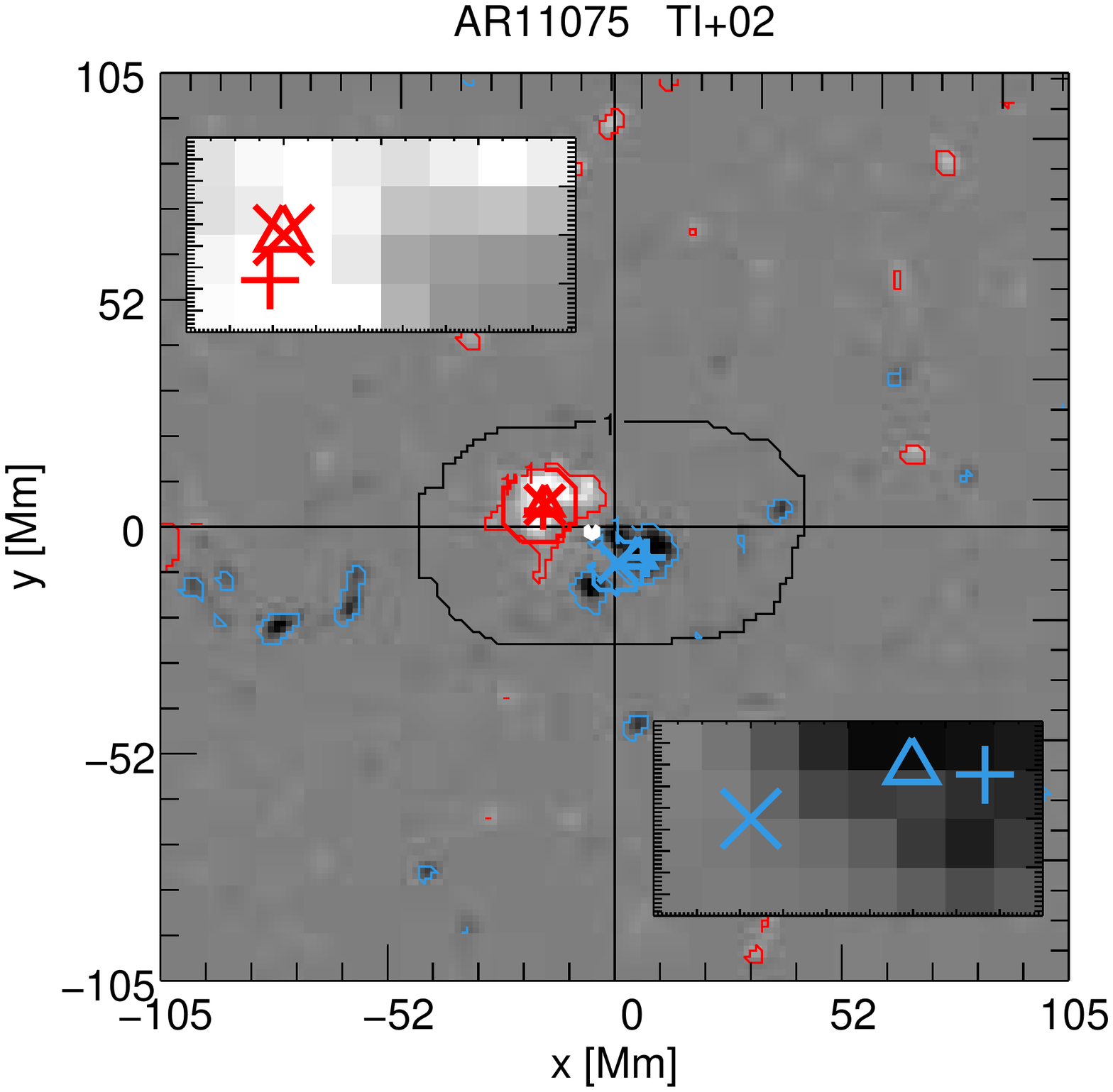}
\includegraphics[width=0.5\textwidth]{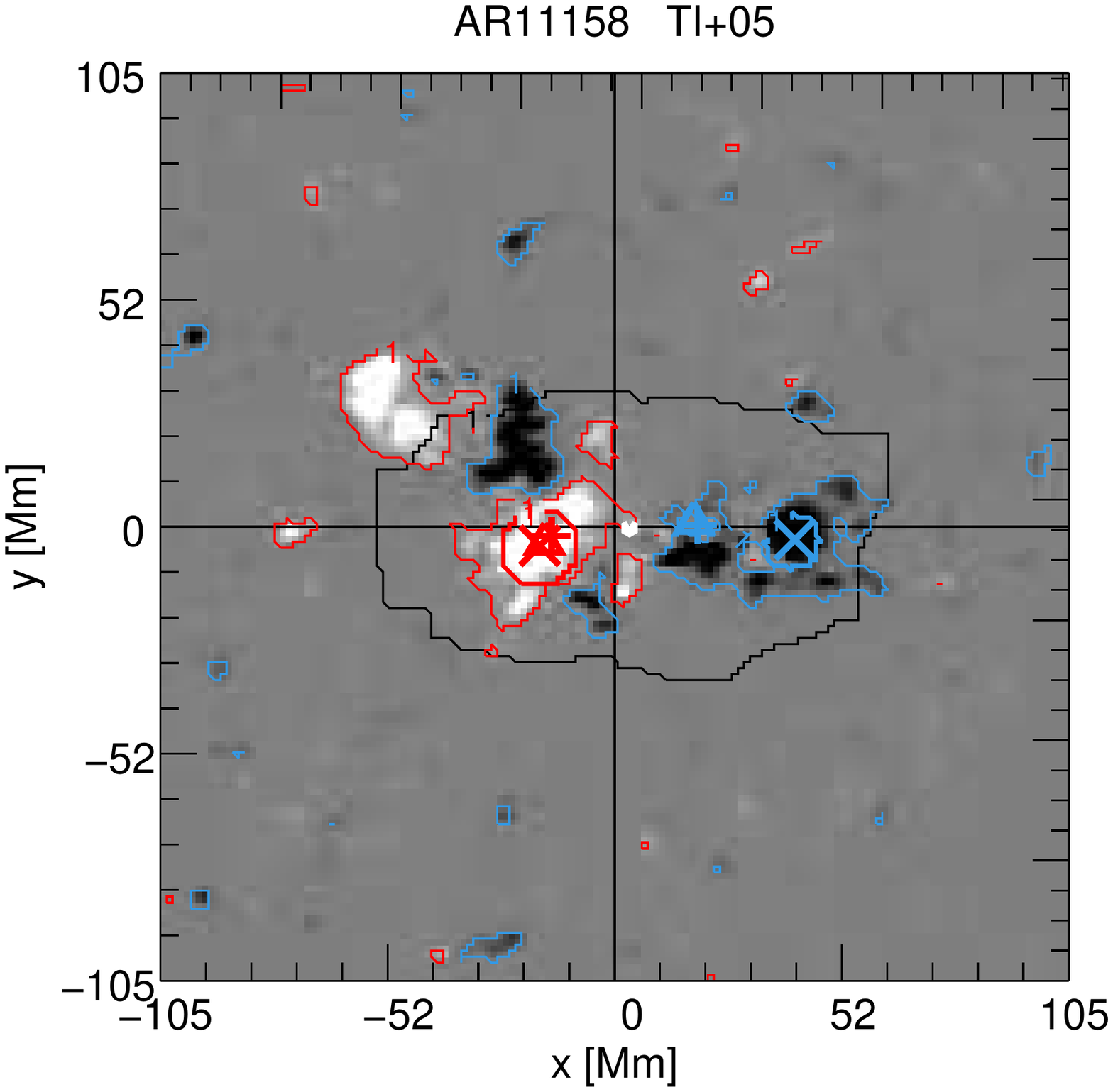}
\caption{
Detected locations of the positive (red) and negative (blue) polarity for two example line-of-sight magnetogram maps (grey-scale saturated at $\pm 200$~G), AR~11075 at \texttt{TI+02} (top) and  AR~11158 at \texttt{TI+05} (bottom).
The crosses indicate the positions detected by the nominal feature detection method.
The triangle indicates the position detected by the flux weighted method.
The  plus  signs indicate the positions  detected by the flux summed method.
The top panel has two insets which show the small differences in the positions more clearly,  by up to a few pixels.
The black contour encloses the search area for all methods.
The red (blue) contour indicates all flux greater than $+20$~G  (less than $-20$~G).
See also Table~\ref{tab:ar11075} and Table~\ref{tab:both}.
}
\label{fig:expos}
\end{figure}

Figure~\ref{fig:expos} shows the location of the identified polarities by three different methods in the line-of-sight magnetic field maps for two active regions.  The three methods we compared were the  feature identification algorithm described above, a flux-weighted method  and a flux-summed  method.  
In the flux-weighted method  we computed the centroid of the line-of-sight magnetic field for each polarity within the search area above (for the following positive polarity) or below (for the leading negative polarity) 20~G defined in Sect.~\ref{sect:method}.
In the flux-summed method we assigned the  line-of-sight magnetic field of positive (negative) polarity above (below) $20$~G ($-20$~G) a value of one and all else zero, and then computed the centroid.  
When the polarities are well isolated and compact, the three methods return results within one pixel of one another (see top panel of Fig.~\ref{fig:expos}).
When the active region is more complex, the results can differ by much more (see bottom panel of Fig.~\ref{fig:expos}). 
This could be due to multiple polarities emerging (as shown in Fig.~\ref{fig:expos}), from polarities splitting up as the active region decays, or from nearby strong field.  
Table~\ref{tab:both} shows the positions  of the negative (leading) and positive (following) polarities detected by the three different methods for two active regions.  
We found the feature identification method described above to be more effective at tracking the location of the polarities associated with the emerging active regions than a flux-weighted centre of gravity method or a flux-summed method.  

\begin{table}
\caption{Positions of the centroids identified by different methods for the leading $(x_l,y_l)$ and following $(x_f,y_f)$ polarities in the AR~11158 line-of-sight magnetic field map for \texttt{TI+05}. See Fig.~\ref{fig:expos}.}
\begin{tabular}{l || c | c | c | c  }
    & $x_f$ & $y_f$ & $x_l$ & $y_l$ \\ 
    & [pix] & [pix] & [pix] & [pix] \\ 
\hline
AR11075 TI+02&  &  &  &  \\
 feature id.   & 244.89 & 260.17 & 257.02 & 250.47 \\ 
 flux weighted    & 244.87 & 260.17 & 260.32 & 251.52 \\ 
 flux summed   & 244.61 & 259.24 & 261.83 & 251.38 \\ 
 \hline 
AR11158 TI+05&  &  &  &  \\
 feature id.   & 244.05 & 253.35 & 286.46 & 254.34 \\ 
 flux weighted    & 244.85 & 253.83 & 269.67 & 257.28 \\ 
 flux summed   & 245.99 & 254.99 & 270.27 & 256.81 \\ 
\end{tabular}

\label{tab:both}
\end{table}

\clearpage
\section{Standard deviation of the standard deviation}\label{app:sd_scatter}

The standard deviation  of the sample standard deviation is 
\begin{equation*}
{\mathrm SD}(s) = \sqrt{ E \left( [E(s)- s]^2 \right) } =  \sqrt{ E(s^2) - E(s)^2 } 
\end{equation*}
where $s^2 = \frac{1}{n-1} \sum_{i=1}^{n} (x_i - \overline{x})^2$,  $x_i$ are the sample data,  $\overline{x}$ is the mean of the sample and $E$ indicates the expectation value. 
Using $E(s^2) = \sigma^2$, where $\sigma$ is the true standard deviation of the entire population we get
\begin{equation*}
{\mathrm SD}(s) = \sqrt{ \sigma^2 - E(s)^2 } ,
\end{equation*}
where  $\sigma^2 =  \frac{1}{n} \sum_{i=1}^{n} (x_i - \mu)^2$, and $\mu$ is the mean of the whole population.
The expected value of the sample is
\begin{equation*}
E(s)
= \sigma \, \sqrt{ \frac{2}{n-1} }  \, \left( \frac{ \Gamma(n/2) }{ \Gamma( \frac{n-1}{2} ) } \right),
\end{equation*}
where $\Gamma (k) = (k-1)!$, and therefore
\begin{equation*}
{\mathrm SD}(s) = \sqrt{ \sigma^2 -  \frac{2 \sigma^2 }{n-1} \, \left( \frac{ \Gamma(n/2) }{ \Gamma( \frac{n-1}{2} ) } \right)^2} .
\end{equation*}


Since we are considering hundreds of active regions in our sample, we used the approximation that $\sigma \approx s$. We also assumed our sample is unbiased, and calculated the standard deviation of the  standard deviation in some measured quantity, $Q$, for $n$ EARs at each time interval as 
\begin{equation}
\mathrm{SD}\Big(~\sigma (Q) ~\Big) 
=  \sqrt{ \sigma(Q)^2 -  \frac{2 \sigma(Q)^2 }{n-1} \, \left( \frac{ \Gamma(n/2) }{ \Gamma( \frac{n-1}{2} ) } \right)^2} 
\label{eqn:sdscat}
\end{equation}
where, for example, $\sigma(Q)= \sigma(\delta)$, the standard deviation in the distance between the polarities. 
This formula describes the calculated uncertainties plotted in Fig.~\ref{fig:bpscat}.

We then made an empirical estimate for comparison. 
To do this, we first computed the standard deviation in some quantity of four subsets of the EARs at each time interval, $\sigma_1(Q), \sigma_2(Q),\sigma_3(Q),\sigma_4(Q)$. 
The standard deviation of the first subset was computed beginning  with the first EAR (ordered by increasing real TAI time of emergence) and every fourth EAR thereafter, the standard deviation of the second subset began with the second EAR and then every fourth EAR thereafter, and so on. 
We then computed the standard deviation of the four subset standard deviations, $\mathrm{SD}\big( ~[\sigma_1(Q),\sigma_2(Q),\sigma_3(Q),\sigma_4(Q)] ~\big)$ at each time interval.

\begin{figure}
\includegraphics[width=0.5\textwidth]{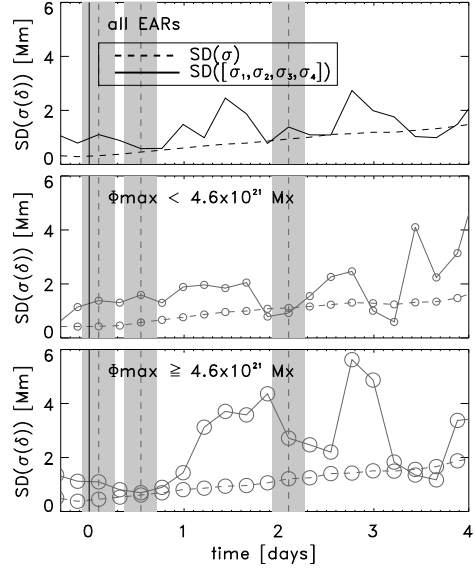}
\caption{
The standard deviation of the standard deviation of the separation between the polarities.
The top panel shows this for all EARs with valid position measurements at each time.
The middle panel shows the same but for EARs with lower maximum flux than the median value, and the bottom panel  for EARs with higher maximum flux than the median value.
}
\label{fig:sd_scatter}
\end{figure}

The comparison of the  calculated, $\mathrm{SD}\big(~\sigma (\delta) ~\big)$, and empirical estimate,  $\mathrm{SD}\big( ~[\sigma_1(\delta),\sigma_2(\delta),\sigma_3(\delta ),\sigma_4(\delta)] ~\big)$, of the standard deviation of the standard deviation in the separation between the polarities in Fig.~\ref{fig:sd_scatter} shows that they are of the same order, however our empirical estimate is noisier and often  larger than the  computed value. 
We did this for each quantity ($\delta$, $\delta x$, $\delta y$, $\gamma$) and for each sample (all EARs, those with low maximum flux and those with high maximum flux). The standard deviations all compared similarly  as that shown in Fig.~\ref{fig:sd_scatter}.

\end{document}